\documentclass[twocolumn,amsmath,amssymb]{revtex4}

\usepackage[colorlinks=true
,urlcolor=blue
,anchorcolor=blue
,citecolor=blue
,filecolor=blue
,linkcolor=blue
,menucolor=blue
,pagecolor=blue
]{hyperref}

\def\[{\left [}
\def\]{\right ]}
\def\({\left (}
\def\){\right )}

\def\Vev#1{\left\langle#1\right\rangle}

\def\ra{\rightarrow}

\def\p{\partial}

\def\a{\alpha}
\def\b{\beta}
\def\d{\delta}

\def\k{\kappa}
\def\l{\lambda}
\def\n{\nabla}
\def\s{\sigma}

\def\D{\Delta}
\def\G{\Gamma}

\def\F{{\cal F}}
\def\J{{\cal J}}
\def\L{{\cal L}}
\def\O{{\cal O}}

\def\W{{\cal W}}

\begin{document}


\title{On AdS/CFT without Massless Gravitons}



\author{Luis Apolo}
\email{lav271@nyu.edu}

\author{Massimo Porrati}
\email{massimo.porrati@nyu.edu}


\affiliation{Center for Cosmology and Particle Physics,\\
Department of Physics, New York University,\\
4 Washington Place, New York, NY 10003, USA}



\begin{abstract}
  We point out that it is possible to define CFT duals to AdS theories with massive gravitons only. This is achieved by considering the product of two or more CFTs on the boundary, each possessing a large-N expansion and a corresponding bulk dual. It is possible to couple the CFTs through a marginal perturbation that makes all but one of the bulk gravitons massive. In the infinite-N limit for one of the CFTs the massless graviton decouples. 
\end{abstract}



\maketitle



  In the AdS/CFT correspondence \cite{Maldacena:1997re,Witten:1998qj,Gubser:1998bc} fields in the bulk act as sources to dual operators on the boundary. In particular, massless gauge fields are dual to conserved currents in the CFT. If the gauge fields are massive their dual operators are not conserved. A remarkable example of this is the correspondence between higher spin theories in AdS$_4$ and 3D $O(N)$ vector models \cite{Klebanov:2002ja,Sezgin:2002rt}. The free $O(N)$ model is dual to Vasiliev's theory \cite{Vasiliev:1992av,Vasiliev:1995dn} with massless higher spin fields of even spin. These fields are dual to the infinite number of conserved currents of the free vector model. In the critical $O(N)$ model \cite{Brezin:1972se,Wilson:1973jj} these currents are no longer conserved. Accordingly, its bulk dual is Vasiliev's theory where the higher spin fields are now massive \cite{Girardello:2002pp,Giombi:2009wh,Giombi:2011ya}. On the other hand the graviton in the AdS/CFT correspondence is massless since its dual, the stress-energy tensor of the boundary theory, is always conserved as a consequence of conformal symmetry.
  
One may think that a CFT dual to an AdS theory with massive gravitons is only possible if it contains an additional massless graviton, but there exist several ways in which one may attempt to modify a pair of AdS/CFT theories to yield duals with only massive gravitons. One possibility is to spontaneously break diffeomorphism invariance and make the graviton massive via quantum corrections from matter fields. This is accomplished by giving the scalars in AdS$_{d+1}$ non-standard, \emph{transparent} boundary conditions characterized by the propagator \cite{Porrati:2001db,Aharony:2006hz}
  \begin{align}
    G(x,x') = a_1\,G_1 (x,x') + a_2\,G_2 (x,x').    \label{eq:propagator}
  \end{align}
  Here $G_k(x,x')$, $k=1,2$, is the propagator corresponding to the standard $D(\D_k,0)$ representation of $SO(d,2)$ where $\sum_k \D_k = d$ and $d/2 > \D_1 > d/2-1$. The $a_k$ are constants satisfying $\sum_k a_k = 1$. Scalars with these boundary conditions can form bound states, vectors in the $D(d+1,1)$ representation of AdS$_{d+1}$, that provide the extra three degrees of freedom necessary for the graviton to become massive. The mass of the graviton must thus be proportional to $a_1 a_2$.
  
  Now recall that a $D(\D,0)$ scalar in AdS$_{d+1}$ with $\D > d/2$ behaves near the boundary as
  \begin{align}
    \phi(z,x) \approx z^{d-\D} \a(x) + z^{\D} \b(x),
  \end{align}
  where $z$ is the radial coordinate of AdS in Poincar\'e coordinates. If we denote by $\W$ a deformation on the CFT and by $\O$ the operator dual to $\phi$, we have \cite{Witten:2001ua}
  \begin{align}
    & \a = \frac{\p \W}{\p \b}, & \b = \frac{1}{2\D - d}\Vev{\O}.
  \end{align}
  In principle it is possible to change the boundary conditions of the bulk scalars, and give a mass to the graviton, by deforming the dual CFT with an appropriate $\W$. This, however, will likely result in a pathological theory at the boundary. For example, in the correspondence between Vasiliev's theory and $O(N)$ vector models the deformation is non-local and breaks conservation of the (boundary) stress-energy tensor. Changing the boundary conditions also introduces problems in the interpretation of the bulk theory since the scalars are now able to cross the boundary of AdS -- it is in this sense that the boundary conditions are transparent.



  Instead, in this Letter we will consider the AdS/CFT duals of Aharony, Clark, Karch, and Kiritsis \cite{Aharony:2006hz,Kiritsis:2006hy}, which contain massive gravitons and have enough parameters so that a limit exists where the massless graviton decouples. This yields a non-trivial duality between a theory in AdS with only massive gravitons and a CFT on the boundary. In \cite{Aharony:2006hz,Kiritsis:2006hy} the product of two or more $d$-dimensional CFTs interacting with one another via a double-trace deformation was studied. Here we consider only the product of two CFTs since their deformation is (almost) unique. Each CFT, denoted by CFT$_i$ with $i=1,2$, lives in the adjoint or vector representations of a group of degree $N_i$ and contains a scalar operator $\O_i$ of dimension $\D_i$ such that $\sum_i\D_i = d$ and $d/2 > \D_1 > d/2-1$. It is possible to have more than one $\O_i$ operator in each CFT and we will consider this case later on.
  
  Each CFT$_i$ has a corresponding dual, a theory in AdS$_{d+1} \times X^{(i)}_{9-d}$ denoted by AdS$^{(i)}$. The AdS theories are not mutually interacting and their geometries are identified at the boundary. This makes transparent boundary conditions feasible. Each theory contains a massless graviton and a scalar $\phi_i$ dual to $\O_i$ in the (standard) $D(\D_i,0)$ representation of $SO(d,2)$. Since $\sum_i \D_i = d$ and $\D_i \ne d/2$ the mass of the scalar fields lies in the range $-d^2/4 < m^2 < -d^2/4 + 1$ which allows for two different solutions corresponding to Dirichlet and Neumann boundary conditions \cite{Breitenlohner:1982bm}. Hence the two AdS theories may be related by a Legendre transformation and correspond to the UV and IR fixed points of two CFTs \cite{Witten:2001ua, Klebanov:1999tb}. The scalar fields are given transparent boundary conditions by deforming the theory at the boundary by
  \begin{align}
    \W = g\,\O_1 \O_2,	\label{eq:deformation}
  \end{align}
  where $g$ is a dimensionless coupling. Here the operators $\O_i$ are normalized so that their two-point functions are independent of $N_i$. With this marginal deformation the boundary (bulk) theories become mutually interacting. In particular, the propagators of the bulk scalars become \cite{Aharony:2005sh}
  \begin{align}
     \Vev{\phi_i\phi_k} = \frac{1}{1+\tilde{g}^2} &\(\begin{array}{rr} 
     G_1 + \tilde{g}^{2} G_2  	&  \tilde{g}\,G_1 - \tilde{g}\,G_2 \\
     \tilde{g}\,G_1 - \tilde{g}\,G_2  & G_2 + \tilde{g}^{2} G_1
     \end{array}\), \label{eq:propagatormatrix}
  \end{align}
where $\tilde{g} = (2\D_1-d)g$. Each propagator in \eqref{eq:propagatormatrix} shares the structure of \eqref{eq:propagator} with $\a_1 \a_2 = \pm \tilde{g}^2/ (1+\tilde{g}^2)^2$ and contributes to the mass of the gravitons. If we denote by $\ell$ the radius of AdS and assume Newton's constant is the same for each AdS$^{(i)}$ theory, $\k^2_i = \k^2 = 8\pi G_N$, the graviton mass matrix is given by
  \begin{align}
    M^2 = \ell^{-d-1}\k^2\s \(\begin{array}{rr}
    1   & -1 \\
    -1  &  1
    \end{array}\), \label{eq:massmatrix}
  \end{align}
and is obtained by isolating the contribution of the $D(d+1,1)$ bound state to the self-energy of the transverse-traceless modes of the gravitons (see \cite{Porrati:2001db,Duff:2004wh} for details). Here $\s$ is a positive number given in \cite{Aharony:2006hz} by
  \begin{align}
    \s = -\frac{\tilde{g}^2}{\(1+\tilde{g}^{2}\)^2}\frac{2(4\pi)^{(3-d)/2}}{(d+2)\G\(\frac{d+3}{2}\)}\prod_i\frac{\D_i\G(\D_i)}{\G\(\D_i-\frac{d}{2}\)}.
  \end{align}

  Let us omit the unnecessary Lorentz indices and denote by $h_i$ the graviton in AdS$^{(i)}$. Diagonalizing \eqref{eq:massmatrix} yields a massless graviton $\tilde{h}_{o} = (1/2)^{1/2}(h_1 + h_2)$, and a massive spin-2 field $\tilde{h}_{m} = (1/2)^{1/2}(h_1 - h_2)$ with mass $m^2 = 2 \ell^{-d-1}\k^2\s$. The graviton is dual to the conserved stress-energy tensor of the boundary CFT, 
  \begin{align}
  T_o = t_1 + t_2 - g\eta \,\O_1\O_2 + \dots, \label{eq:tmunu}
  \end{align}
  where we have omitted improvement terms that make $T_o$ traceless. Here $\eta$ is the AdS background metric and $t_i$ is the stress-energy tensor of CFT$_i$. On the other hand the massive spin-2 is dual to a linear combination of $t_i$ and $g\eta\,\O_1\O_2$ that is not conserved and orthogonal to $T_o$.


  
  This summarizes the construction of \cite{Aharony:2006hz,Kiritsis:2006hy} in the simplest case. We now show that it is possible to obtain AdS/CFT duals with only massive gravitons by letting $\k_1 \ne \k_2$ and taking a limit where either $\k_1$ or $\k_2$ vanishes. To see that the massless graviton decouples let us consider the quadratic Lagrangian for the gravitons in the bulk\footnote{Note that there maybe corrections to the kinetic terms that render the corresponding matrix non-diagonal. These corrections are subleading in $1/N$ so they may be ignored.}
  \begin{equation}
  \begin{split}
    \L_2  =  - \frac{1}{2}\(h_1 \;\; h_2 \) & \(\begin{array}{cc} \k_1^{-2} \square & 0 \\ 0 & \k_2^{-2}\square \end{array}\) \(\begin{array}{c} h_1 \\ h_2 \end{array}\) \\
    - & \frac{\sigma}{2}  \(h_1 \;\; h_2 \)\(\begin{array}{rr} 1 & -1 \\ -1 & 1 \end{array}\) \(\begin{array}{c} h_1 \\ h_2\end{array}\), \label{eq:L}
  \end{split} 
  \end{equation}
  where the mass term is given by \eqref{eq:massmatrix} without the factor of $\k^{2}$ and we have set $\ell = 1$. Here we have greatly simplified the notation so that $h_{(i)} \square h_{(i)}$ stands for the quadratic Lagrangian of a massless graviton in AdS, namely
  \begin{align}
    h_{(i)}^{\mu\nu}(-2\n^\a\n_\mu\,\d^\b_\nu + \n^2\,\d^\a_\mu\,\d^\b_\nu + \dots) h_{\a\b(i)}, 
  \end{align}
  with $\n_{\a}$ the background covariant derivative; while $\s\,h_{(i)}\,h_{(k)}$ stands for the Fierz and Pauli \cite{Fierz:1939ix} mass term
  \begin{align}
 h_{\mu\nu(i)}h^{\mu\nu}_{(k)}-h^{\mu}_{\mu(i)}h^{\nu}_{\nu(k)} 
  \end{align}

  In terms of the canonically normalized field $H_i = \k_i^{-1}h_i$ the Lagrangian becomes 
  \begin{equation}
  \begin{split}
    \L_2 =  - & \frac{1}{2}\(H_1 \;\; H_2 \)\(\begin{array}{cc} \square & 0 \\ 0 & \square \end{array}\)\(\begin{array}{c} H_1 \\ H_2\end{array}\)  \\
    - & \frac{\s}{2}\(H_1 \;\; H_2 \)\(\begin{array}{cc} \k_1^2 & -\k_1\k_2 \\ -\k_1\k_2 & \k_2^2 \end{array}\)\(\begin{array}{c} H_1 \\ H_2\end{array}\). \label{eq:canonicalL}
  \end{split}
  \end{equation}
  Diagonalizing the mass term then yields
  \begin{align}
    \L_2 = - \frac{\k_1^2 + \k_2^2}{4\k_1^2\k_2^2}\,\tilde{h}_o \square \tilde{h}_o - \frac{1}{\k_1^{2} + \k_2^{2}} \,\tilde{h}_m \square \tilde{h}_m  - \s \tilde{h}_m^2 \label{eq:diagonalL}
  \end{align}
  where $\tilde{h}_o$ and $\tilde{h}_m$ are given by
  \begin{align}
    \tilde{h}_o & \equiv \frac{\sqrt{2}}{\k_1^2 + \k_2^2} \( \k_2^2\,h_1 + \k_1^2\,h_2 \) \\
    \tilde{h}_m & \equiv \frac{1}{\sqrt{2}}\( h_1 - h_2 \).
  \end{align}

  It is clear from \eqref{eq:diagonalL} that for $\k_1 = \k_2$ there is no limit in which only $\tilde{h}_o$ decouples. With $\k_1 \ne \k_2$ taking the $\k_i \ra 0$ limit for either value of $i$, say $i = 1$, leads to a theory with only a massive graviton. Its mass is now given by $m^2 = \k_2^2\s$ and is half of that of the theory with $\k_1 = \k_2$. Now recall that in the AdS/CFT correspondence Newton's constant is related to $N_i$ by 
  \begin{align}
    G_N^{(i)} \propto \ell^{d-1}(1/N_i)^p,
  \end{align}
  where $p = 1\,(2)$ for a dual CFT living in the vector (adjoint) representation. Taking the $\k_i \rightarrow 0$ limit in the bulk corresponds to taking the $N_i\rightarrow \infty$ limit in the boundary. Thus the dual theory at the boundary is given by the product of CFTs with deformation \eqref{eq:deformation} where $N_1 \ra \infty$ and $N_2$ is large but finite.

  The decoupling of the massless graviton in the bulk is equivalent to the vanishing of all correlators, other than the two-point function, of the boundary stress-energy tensor. To see this recall that in the AdS/CFT correspondence the on-shell partition function in the bulk is the generating functional of operators in the dual CFT. In particular, two-point functions are determined by the quadratic action. Let us denote by $T_o$ and $T_m$ the operators dual to $\tilde{h}_o$ and $\tilde{h}_m$, respectively. Since $\k^2_i \propto \ell^{d-1}(1/N_i)^p$, inspection of \eqref{eq:diagonalL} yields two-point functions with the following dependence on $N_i$ when $N_i$ is large
  \begin{align}
    & \Vev{T_o\,T_o} \simeq N_1^p + N_2^p,  & \Vev{T_m\,T_m} & \simeq \frac{(N_1 N_2)^p}{N_1^p + N_2^p}.
  \end{align}
  This can be verified by inspection of the two-point functions using \eqref{eq:tmunu} and the definition of $T_m$ given by eq. (2.9) of \cite{Aharony:2006hz}, with the latter normalized by $(N_1^p + N_2^p)^{1/2}$. 

  Normalizing $T_o$ and $T_m$ so that their two-point functions are independent of $N_i$ yields, in the limit $N_1 \ra \infty$,
  \begin{align}
    & T_o \ra (1/N_1)^{p/2}\,T_o,  & T_m \ra (1/N_2)^{p/2}\,T_m.
  \end{align}
  With all other operators $\J_i \in$ CFT$_i$ normalized in a similar fashion, 
  \begin{align}
    & \J_i \ra (1/N_i)^{p/2}\J_i, & \O_i \ra (1/N_i)^{p/2}\O_i, 
  \end{align}
  the $n$-point functions of $T_o$ scale with $N_1$ as follows
  \begin{align}
    \big\langle T_o\,\F[T_o,T_m,\J_i,\O_i]\big\rangle \simeq \frac{1}{N_1^q},
  \end{align}  
  where $\F$ is any polynomial of degree $n-1$ and $q > 0$ for $n > 2$. Hence the $n$-point functions of $T_o$ vanish in the $N_1 \ra \infty$ limit. On the other hand correlation functions of $T_m$ are finite with all terms of $\O(g^3)$ vanishing.

  It is not difficult to find and example where this construction is applicable. Consider again the correspondence between higher spin theories in AdS$_4$ and $O(N)$ vector models on the boundary. As mentioned earlier there are two faces to the correspondence. On the one hand we have Vasiliev's theory with massless gauge fields of even spin on the bulk and the free vector model on the boundary. This theory contains a scalar in the $D(1,0)$ representation of $SO(3,2)$. Accordingly its boundary dual, $\O_1$, has dimension $\Delta_1 = 1$. On the other hand the scalar in Vasiliev's theory with massive gauge fields lives in the $D(2,0)$ representation of AdS$_4$; its dual operator $\O_2$ has dimension $\Delta_2 = 2 + \O(1/N_2)$. Since $\Delta_1 + \Delta_2 = 3$, the product of the two vector models with deformation \eqref{eq:deformation} makes one of the bulk gravitons massive. By sending $N_1 \ra \infty$ with $N_2$ large but finite we obtain an explicit example of a correspondence with only massive gravitons in the bulk.



  In this Letter we have considered the simplest deformation of the product of two CFTs. In general there may be $\rho$ operators in CFT$_{i}$ of dimension $\Delta_i$ denoted by $\O^{(a)}_i$. In this case eq. \eqref{eq:deformation} may be generalized to
  \begin{align}
    \W = \sum_{a,b}^\rho g_{ab}\, \O^{(a)}_1\O^{(b)}_2. \label{eq:newdef}
  \end{align}
  If for simplicity we take $g_{ab}$ to be symmetric it may be diagonalized so that $\W$ becomes
  \begin{align}
    \W = \sum_{a}^\rho \l_a \, \widetilde{\O}^{(a)}_1 \widetilde{\O}^{(a)}_2,
  \end{align}
  where $\l_a$ are the eigenvalues of $g_{ab}$ and $\widetilde{\O}_i^{(a)}$ is a linear combination of $\O^{(a)}_i$. The pair of operators $\{\widetilde{\O}_1^{(a)}, \widetilde{\O}_2^{(a)}\}$ changes the boundary conditions of the corresponding pair of dual scalars. Thus each pair of scalars contributes to the mass of the graviton and their propagators are given by \eqref{eq:propagatormatrix} with $\tilde{g} \ra \tilde{\l}_a = (2\Delta_1 -d)\l_a$.  With $N_1 \ra \infty$ and $N_2$ large but finite, the mass of the graviton is given by $m^2 = \k^2_2\, \tilde{\s}$ where
  \begin{align}
    \tilde{\s} = \sum_a \frac{-\tilde{\l}_a^2}{(1 + \tilde{\l}_a^{2})^2} \frac{2(4\pi)^{(3-d)/2}}{(d+2)\G\(\frac{d+3}{2}\)}\prod_i\frac{\D_i\G(\D_i)}{\G\(\D_i-\frac{d}{2}\)}, \label{eq:newmass}
  \end{align}
  and $\tilde{\l}_a^2/(1 + \tilde{\l}_a^{2})^2$ has a maximum at $\tilde{\l}_a = \pm 1$. 

Generically, for $g_{ab} \sim \O(1)$ deformation \eqref{eq:newdef} leads to a graviton mass smaller than that induced by \eqref{eq:deformation}. It is possible to choose $g_{ab}$ such that $\tilde{\l}_a \sim \O(1)$. In this case \eqref{eq:newmass} leads to an enhancement of the graviton mass by a factor of $\rho$. On the other hand the presence of $\rho$ species of scalars in the bulk will contribute to the wave function renormalization of the gravitons. This will reduce Newton's constant $G^{(i)}_N$ by a factor of $\rho^{-1}$ \cite{Dvali:2007hz}. Thus the more general case given in \eqref{eq:newdef} may reduce the mass of the graviton but leaves the main result unchanged.



  We have seen that it is possible to have a correspondence between a theory in AdS with only massive gravitons and a CFT on the boundary. The construction of \cite{Aharony:2006hz,Kiritsis:2006hy} was necessary so that (1) the field dual to the massive graviton is some operator other than the conserved stress-energy tensor of the CFT; and (2) there is a limit in which the massless graviton decouples. Using the methods of refs~\cite{Giombi:2009wh,Giombi:2010vg,Giombi:2011ya} it should be possible to check explicitly whether Vasiliev's theory with massive gravitons is dual to the product of vector models with deformation \eqref{eq:deformation} where we take either $N_1$ or $N_2$ to infinity.

\bigskip


\begin{acknowledgments}
We would like to thank Sergei Dubovsky and Matthew Kleban for useful comments. M.P. is supported in part by NSF grant PHY-0758032, and by ERC Advanced Investigator Grant No.\@ 226455 {\em Supersymmetry, Quantum Gravity and Gauge Fields (Superfields)}.
\end{acknowledgments}




\end{document}